# A Novel Quantum Fourier Ordinary Differential Equation Solver for Solving Linear and Nonlinear Partial Differential Equations

Y. Xiao[1], L.M. Yang[1, 2, 3, *], C. Shu[4, †], Y. J. Du and Y. X. Song

[1]Department of Aerodynamics, College of Aerospace Engineering, Nanjing University of Aeronautics and Astronautics, Nanjing 210016, China

[2]State Key Laboratory of Mechanics and Control for Aerospace Structures, Nanjing University of Aeronautics and Astronautics, Nanjing 210016, China

[3]MIIT Key Laboratory of Unsteady Aerodynamics and Flow Control, Nanjing University of Aeronautics and Astronautics, Nanjing 210016, China

[4]Department of Mechanical Engineering, National University of Singapore, Singapore 117576, Singapore

## Abstract

In this work, a novel quantum Fourier ordinary differential equation (ODE) solver is proposed to solve both linear and nonlinear partial differential equations (PDEs). Traditional quantum ODE solvers transform a PDE into an ODE system via spatial discretization and then integrate it, thereby converting the task of solving the PDE into computing the integral for the driving function $f(x)$. These solvers rely on the quantum amplitude estimation algorithm, which requires the driving function $f(x)$ to be within the range of [0, 1] and necessitates the construction of a quantum circuit for the oracle $\mathcal{R}$ that encodes $f(x)$. This construction can be highly complex, even for simple functions like $f(x) = x$. An important exception arises for the specific case of $f(x) = \sin^2(mx+c)$, which can be encoded more efficiently using a set of $R_y$ rotation gates. To address these challenges, we expand the driving function $f(x)$ as a Fourier series and propose the Quantum Fourier ODE Solver. This approach not only simplifies the construction of the oracle $\mathcal{R}$ but also removes the restriction that $f(x)$ must lie within [0,1]. The proposed method was evaluated by solving several representative linear and nonlinear PDEs, including the Navier-Stokes (N-S) equations. The results show that the quantum Fourier ODE solver produces results that closely match both analytical and reference solutions.

---

[*]Corresponding author, E-mail: lmyang@nuaa.edu.cn.
[†]Corresponding author, E-mail: mpeshuc@nus.edu.sg.

# 1 Introduction

The advent of classical computers in the last century led to the emergence of computational fluid dynamics (CFD) [1], driving its rapid development and widespread applications through CFD algorithms deployed on large parallel computers [2]. However, CFD remains a computationally intensive task, with its computational complexity growing significantly as the mesh size increases [3-4]. As Moore's Law slows, the performance gains of classical computers face mounting challenges [5]. While multi-core and heterogeneous computing continues to evolve [6], no transformative hardware breakthroughs have yet materialized. Fortunately, the advent and breakthrough of quantum computing (QC) offers a potential new paradigm for CFD [7-8]. Unlike classical computers, quantum computers leverage the fundamental principles of quantum mechanics (such as superposition and entanglement) for computation, offering a competitive edge in addressing certain complex problems [9-10].

Quantum computers possess the unique feature of superposition. Unlike classical bits, which exist strictly as 0 or 1, quantum bits (qubits) can exist in a superposition of both 0 and 1 states. As a result, $n$ qubits can represent a linear combination of $2^n$ states, whereas a classical $n$-bit system can represent only one of $2^n$ possible states. This feature underpins the development of a wide array of quantum algorithms, including Shor's [11] and Grover's [12] algorithms. Shor's and Grover's algorithms are respectively quantum algorithms for integer factorization and unstructured database search, and they have been shown to offer significant speedup over classical algorithms. In particular, Shor's algorithm is considered a direct threat to the traditional Rivest–Shamir–Adleman (RSA) encryption scheme [13]. In the field of scientific computing, Harrow, Hassidim, and Lloyd [14] proposed a quantum algorithm for solving linear systems of equations, known as the Harrow-Hassidim-Lloyd (HHL) algorithm, which demonstrates exponential speedup over classical methods for solving certain linear systems. Since CFD solvers typically discretize governing equations on meshes to form a system of linear equations, transforming the task of solving partial differential equations (PDEs) into solving systems of linear equations, the HHL algorithm has been incorporated as a subroutine in CFD solvers [15-16]. However, the HHL algorithm relies on the quantum Fourier transform (QFT) [17] and quantum phase estimation algorithm (QPEA) [18], requiring substantial quantum circuit resources, which renders its unsuitability for current noisy intermediate-scale quantum (NISQ) devices [19]. To address this issue, Bravo-Prieto et al. [20] developed the variational quantum linear solver (VQLS), which is a variational quantum algorithm for solving linear systems. Similarly, VQLS has been employed as a subroutine in CFD solvers, and successfully applied to problems such as the Poisson equation [21-22], heat conduction equation [23], and potential flow [24]. VQLS is essentially a hybrid classical-quantum algorithm that requires fewer quantum circuit resources compared to the HHL algorithm, making it more practical for current NISQ devices [25-26].

Despite quantum computers still being in the early stages of development, the potential benefits of quantum speedup are becoming increasingly evident [27]. Moreover, the superposition property of quantum computers offers another potential advantage in computational memory. For instance, conducting a CFD simulation of a real-world aircraft flight, where the Reynolds number (Re) is approximately $10^8$, would require around $Re^{9/4} \approx 10^{18}$ mesh points [28]. If each mesh point stores a single flow variable with 64-bit precision, the memory requirement would be approximately $10^{18} \times 64$ bits, equivalent to 8,000 petabytes ($8 \times 10^9$ GB) [27]. However, by using amplitude encoding of classical information on a quantum computer, the number of qubits required would be approximately 60 (~15/2log(Re)), which is well within the capabilities of current quantum hardware [29]. Furthermore, the lattice Boltzmann method (LBM) [30-31], a mesoscopic approach, generally requires more computational storage than traditional CFD methods like finite difference [32] or finite volume techniques [33]. To highlight the advantages of quantum computing based on LBM, several attempts have been made. For instance, Budinski [34] developed a quantum version of LBM (QLBM) to solve the convection-diffusion equation. Additionally, Kocherla et al. [35] introduced a two-circuit QLBM for solving the Navier-Stokes (N-S) equations, significantly reducing the quantum circuit resources required as compared to traditional QLBM circuits.

Despite these advancements, the quantum LBM faces challenges when dealing with nonlinear collision terms, which often require linearization [28, 36]. Similarly, both the HHL algorithm and VQLS method necessitate discretizing PDEs into linear equation systems, which is typically difficult for nonlinear PDEs [37]. Recently, Gaitan [38-39] proposed a method for discretizing nonlinear PDEs into ordinary differential equations (ODEs) and solving them using a quantum ODE solver. Initially introduced by Kacewicz [40], the quantum ODE solver transforms the task of solving ODEs into computing integrals of the driving function $f(x)$ on the right-hand side of the ODE. Kacewicz employed Novak's quantum integration algorithm (QIA) [41] for this purpose, which is the only step in the entire process requiring a quantum computer. However, the QIA is based on the quantum amplitude estimation algorithm (QAEA) [42], which restricts the range of the driving function $f(x)$ to [0,1]. Furthermore, the QIA requires constructing an oracle $\mathcal{R}$ to encode the driving function, which was noted by Herbert [43] as a quite complex task, even for simple functions.

A notable exception occurs for integrals of the form $f = \sin^2(mx + c)$, where the oracle $\mathcal{R}$ consists solely of a series of $R_y$ gates. Building on this observation, Herbert [43] proposed a Fourier quantum Monte Carlo integration (QMCI) algorithm for computing integrals, in which the integrand is expanded into a Fourier series, thereby converting the original integral into the computation of trigonometric function integrals. The Fourier QMCI algorithm not only retains the quadratic speedup over classical integration methods but also eliminates the restriction of $f(x)$ being confined to the range of [0,1]. In this work, we will incorporate Herbert's Fourier QMCI algorithm into the quantum ODE solver and propose a novel quantum Fourier ODE solver for solving both linear and nonlinear PDEs. Specifically, we first spatially discretize the PDE to form an ODE system $u_t = f(u)$, then integrate this ODE system using a standard time-stepping scheme $u(t_{i+1}) = u(t_i) + \int_{t_i}^{t_{i+1}} f(u(t)) dt$ Finally, the integrand $f(x)$ is expanded into a Fourier series, transforming the original integral into the trigonometric function-type integral, which is then computed using the QAEA.

The structure of this paper is arranged as follows: Section 1 provides an introduction; Section 2 details the fundamentals of both the traditional quantum ODE solver and quantum Fourier ODE solver; Section 3 applies the developed method to solve several test examples; Section 4 gives a brief summary.

## 2 Quantum ODE Solvers

### 2.1 Quantum integration algorithm (QIA)

In this section, we begin by presenting Novak's QIA algorithm, which leverages QAEA to compute integrals. QAEA is essentially a combination of the quantum amplitude amplification algorithm (QAAA) [42] and the QPEA [18]. The QAAA is an extension of Grover's search algorithm, retaining the quadratic quantum speedup offered by Grover's algorithm. Fig. 1 illustrates the quantum circuit of the QAAA. Suppose that we have a unitary operator $\mathcal{A}$ acting on $n+1$ qubits, yielding the following state

$$|\Psi\rangle = \mathcal{A}|0\rangle_n|0\rangle = \sqrt{1-a}|\psi_0\rangle_n|0\rangle + \sqrt{a}|\psi_1\rangle_n|1\rangle, \tag{1}$$

where $a \in [0, 1]$, and $|\psi_1\rangle_n$ and $|\psi_0\rangle_n$ represent the good and bad states, respectively. The primary objective of QAEA is to estimate $a$, which corresponds to the probability of measuring the good state. Since a single application of $\mathcal{A}$ may result in a low probability of the good state, QAAA is employed to amplify this probability. The amplification is achieved using the following unitary operator

$$Q = -\mathcal{A}S_0\mathcal{A}^{-1}S_\chi, \tag{2}$$

where $\mathcal{A}^{-1}$ represents the inverse of $\mathcal{A}$, the operator $S_0$ marks the state $|0\rangle_{n+1}$ with a negative sign (leaving other states unaffected), while the operator $S_\chi$ modifies the sign for the good state $|\psi_1\rangle_n$, i.e., $S_\chi|\psi_1\rangle_n|1\rangle = -|\psi_1\rangle_n|1\rangle$. Since $0 \leq a \leq 1$, we define a parameter $\theta \in [0, \pi/2]$ such that $\sin^2\theta = a$, allowing Eq. (1) to be rewritten as

$$|\Psi\rangle = \mathcal{A}|0\rangle_n|0\rangle = \cos\theta|\psi_0\rangle_n|0\rangle + \sin\theta|\psi_1\rangle_n|1\rangle. \qquad (3)$$

By repeatedly applying the operator $Q$ from Eq. (2) $m$ times on $|\Psi\rangle$, the resulting state is

$$Q^m|\Psi\rangle = \cos((2m+1)\theta)|\psi_0\rangle_n|0\rangle + \sin((2m+1)\theta)|\psi_1\rangle_n|1\rangle. \qquad (4)$$

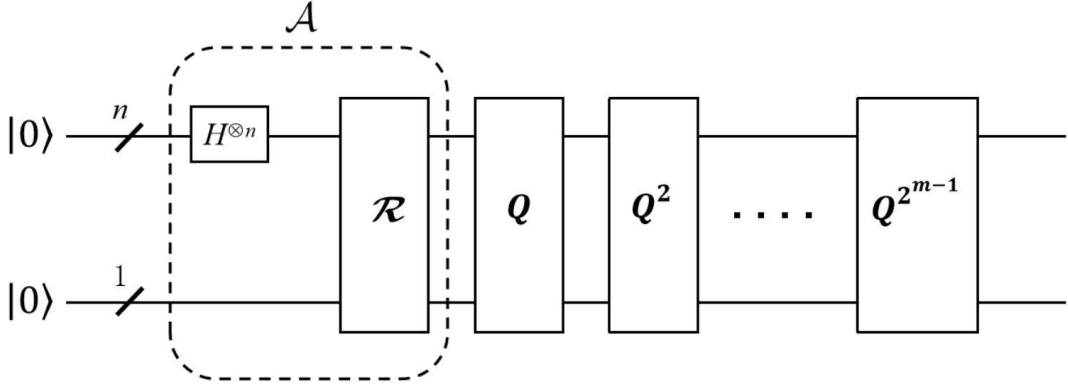

**Fig. 1** Quantum circuit for QAAA.

When $\theta$ is sufficiently small such that $(2m+1)\theta < \pi/2$, apply $Q^m$ to $\mathcal{A}|0\rangle_n|0\rangle$ and then measure results in a quadratically larger probability of obtaining the good state than measuring $\mathcal{A}|0\rangle_n|0\rangle$ directly. The traditional QAEA employs QPEA to estimate $\theta$, which necessitates a quantum circuit that implements multiple controlled-$Q$ operations and collects the amplitudes via the inverse quantum Fourier transform (QFT). The circuit of QAEA is illustrated in Fig. 2.

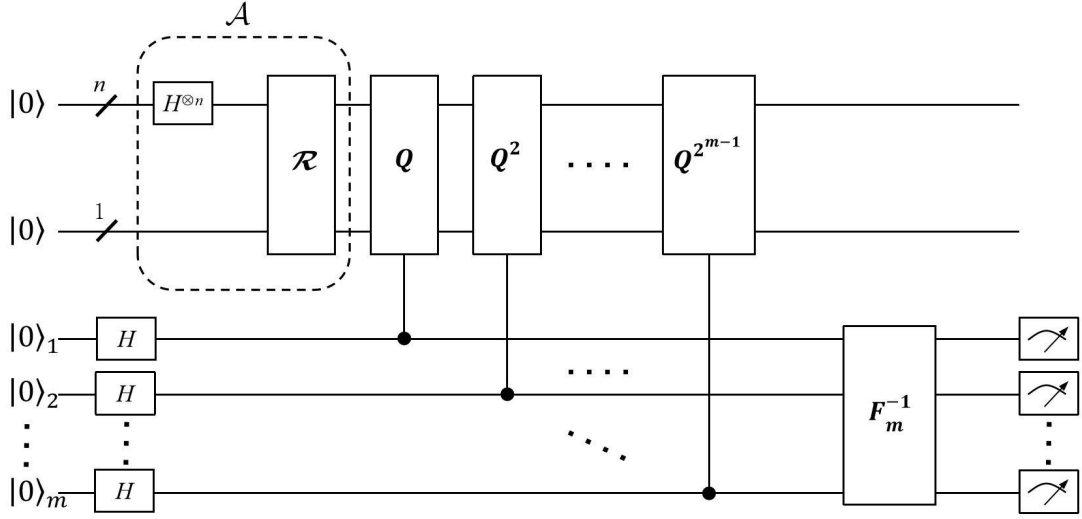

**Fig. 2** Quantum circuit for QAEA. $F_m^{-1}$ represents the inverse QFT of $m$ qubits.

Next, we discuss how to use QAEA to compute integrals. Consider the following integral

$$I = \int_0^1 f(z)dz. \tag{5}$$

Since $0 \leq a \leq 1$ in Eq. (1), the function $f(z)$ needs to be scaled to ensure that the integrand falls within the range [0, 1]. Suppose that $f(z)$ is a continuous function and introduce the following transformation

$$g(z) = \frac{f(z) - f_{min}}{f_{max} - f_{min}}, \tag{6}$$

where $0 \leq g(z) \leq 1$, and $f_{max}$ and $f_{min}$ are the maximum and minimum values of $f(z)$, respectively. Therefore, Eq. (5) can be rewritten as

$$I = f_{min} + (f_{max} - f_{min})\mathcal{G}, \tag{7}$$

where

$$\mathcal{G} = \int_0^1 g(z)dz. \tag{8}$$

The definite integral $\mathcal{G}$ possesses the properties required for applying QAEA. To use

QAEA to compute an approximation of $\mathcal{G}$, we can replace $\mathcal{G}$ by its Riemann sum $\mathcal{G}_R$

$$\mathcal{G}_R = \frac{1}{N}\sum_{i=0}^{N-1} g(z_i). \tag{9}$$

As shown in Fig. 2, the operator $\mathcal{A}$ in QAEA is composed of two components

$$H^{\otimes n}|0\rangle_n = \frac{1}{\sqrt{N}}\sum_{i=0}^{N-1}|i\rangle_n, \tag{10}$$

where $H^{\otimes n}$ denotes the Hadamard gate $H$ applied to the qubits in the first register, and

$$\mathcal{R}|i\rangle_n|0\rangle = |i\rangle_n\left(\sqrt{1-g(z_i)}|0\rangle + \sqrt{g(z_i)}|1\rangle\right), \tag{11}$$

which encodes the function $g(z)$ on an ancillary qubit added to the circuit. Therefore, applying the operator $\mathcal{A}$ to the $(n + 1)$-qubit initial state gives

$$|\psi\rangle = \mathcal{A}|0\rangle_{n+1} = \mathcal{R}\left(H^{\otimes n} \otimes \mathbf{I}\right)|0\rangle_{n+1} = \sum_{i=0}^{N-1}\frac{1}{\sqrt{N}}|i\rangle_n\left(\sqrt{1-g(z_i)}|0\rangle + \sqrt{g(z_i)}|1\rangle\right), \tag{12}$$

where $\mathbf{I}$ represents the identity operator. Now, we introduce the normalized and mutually orthogonal states $|\psi_0\rangle$ and $|\psi_1\rangle$,

$$\begin{aligned}|\psi_0\rangle_n &= \frac{1}{\sqrt{N}}\sum_{i=0}^{N-1}\sqrt{\frac{1-g(z_i)}{1-\bar{g}}}|i\rangle_n, \\ |\psi_1\rangle_n &= \frac{1}{\sqrt{N}}\sum_{i=0}^{N-1}\sqrt{\frac{g(z_i)}{\bar{g}}}|i\rangle_n,\end{aligned} \tag{13}$$

and we can rewrite the state $|\psi\rangle$ as

$$|\psi\rangle = \sqrt{1-a}|\psi_0\rangle_n|0\rangle + \sqrt{a}|\psi_1\rangle_n|1\rangle. \tag{14}$$

This is structurally similar to Eq. (1), and

$$\bar{g} = \frac{1}{N}\sum_{i=0}^{N-1}g(z_i) = a. \tag{15}$$

Therefore, the computation of the integral $\mathcal{G}$ can now be viewed as an estimate of the amplitude $a$. To achieve this, we introduce a Grover-like operator $\mathbf{Q} = U_\psi U_{\psi 0}$, defined as

$$\begin{aligned} \boldsymbol{Q} &= U_\psi U_{\psi 0}, \\ U_{\psi 0} &= \mathbf{I}^{\otimes n+1} - 2\mathbf{I}^{\otimes n}|0\rangle\langle 0|, \\ U_\psi &= \mathbf{I}^{\otimes n+1} - 2|\psi\rangle\langle\psi|, \end{aligned} \qquad (16)$$

Overall, by defining $a = \sin^2\theta$, we can then use Eq. (14) and the QAEA algorithm described above to compute the integral $\mathcal{G}$.

## 2.2 Traditional quantum ODE solver

Consider the following general form of the differential equation

$$u_t + \mathcal{L}_x[u] = 0, \qquad \boldsymbol{x} \in \Omega \in R^d, t \in [0,T] \qquad (17)$$

where $\mathcal{L}_x[\cdot]$ is the general differential operator. The spatial discretization is used in Eq. (17), and transforming it into a set of nonlinear ordinary differential equations (ODEs)

$$u_t = f(u). \qquad (18)$$

By integrating the above equation, we obtain

$$u(T) = u(0) + \int_0^T f(u(\tau))d\tau. \qquad (19)$$

Following Kacewicz's quantum ODE solver [40], the time interval $0 \leq t \leq T$ is first partitioned into $n$ subintervals $I_i = [t_i, t_{i+1}]$ with a duration of $h = T/n$, where $\{t_0 = 0, \ldots, t_n = T\}$. Therefore, a standard time stepping scheme yields

$$u(t_{i+1}) = u(t_i) + \int_{t_i}^{t_{i+1}} f(u(\tau))d\tau. \qquad (20)$$

For all subintervals, $n$ parameters $\{y_i|\ 0 \leq i \leq n-1\}$ are introduced, with $y_i$ corresponding to the initial time $t_i$ of the $i$-th subinterval $I_i$. Specifically, the parameters $\{y_i|\ 1 \leq i \leq n-1\}$ approximate the exact solution at time $t_i$

$$y_i \approx u(t_i), \tag{21}$$

Meanwhile, $y_0$ is specified as the initial condition of the ODE

$$y_0 = u(0). \tag{22}$$

Then, each subinterval $I_i$ is further divided into $N_k = n^{k-1}$ sub-subintervals $I_{i,j} = [t_{i,j}, t_{i,j+1}]$ with a duration of $\bar{h} = h/N_k = T/n^k$, where $\{t_{i,0} = t_i, \cdots, t_{i,N_k} = t_{i+1}\}$. The approach for selecting $n$ and $k$ will be detailed in Section 3. Let $A_i(t)$ represent the approximate solution of the exact solution $u(t)$ in the subinterval $I_i$, and denote the approximate solution in the sub-subinterval $I_{i,j}$ as $A_{i,j}(t)$. The Taylor series expansion of $A_{i,j}(t)$ at $t_{i,j}$ is given by

$$A_{i,j}(t) = A_{i,j}(t_{i,j}) + \sum_{k=0}^{r} \frac{1}{(k+1)!} \frac{d^k f}{dt^k}\bigg|_{A_{i,j}(t_{i,j})} (t - t_{i,j})^{k+1} + O(\bar{h}^{r+2}) \tag{23}$$

The parameter $r$ is chosen so that the error $O(\bar{h}^{r+2})$ is sufficiently small. Since the approximate solution $A_i(t)$ is required to be continuous throughout $I_i$, it must be equal at the common boundary of two adjacent sub-subintervals

$$A_{i,j}(t_{i,j+1}) = A_{i,j+1}(t_{i,j+1}). \tag{24}$$

Additionally, $A_i(t)$ is required to take the value $y_i$ at the initial time $t_i$ of the $i$-th subinterval $I_i = [t_i, t_{i+1}]$

$$A_i(t_i) = A_{i,0}(t_i, 0) = y_i. \tag{25}$$

Therefore, once the parameters $\{y_i | 0 \leq i \leq n-1\}$ are determined, the approximate solution $A_i(t)$ can be constructed using Eqs. (23)-(25). Specifically, starting with $y_0$, the initial value $A_{0,0}(t_{0,0})$ of the approximate solution $A_{0,0}(t)$ for the first sub-subinterval $I_{0,0}$ of the first subinterval $I_0$ is obtained using Eq. (25). The approximate solution $A_{0,0}(t)$ for the entire sub-subinterval $I_{0,0}$ is then computed using

Eq. (23), and the initial value $A_{0,1}(t_{0,1})$ for the next sub-subinterval $I_{0,1}$ is determined using Eq. (24). By repeating this process for each sub-subinterval $I_{0,j}$ of $I_0$, the approximate solutions $A_{0,i}(t)$ for all sub-subintervals are obtained, which leads to the approximate solution $A_0(t) = \bigcup_{j=0}^{N_k-1} A_{0,j}(t)$ for the subinterval $I_0$. Finally, this process is repeated for all subintervals $I_i$, yielding the approximate solutions $A_i(t)$ for each subintervals, and ultimately constructing the approximate solution $A(t) = \bigcup_{i=0}^{n-1} A_i(t)$ for the entire interval $I$.

In the above process, the calculation of $\{y_i|\ 0 \leq i \leq n-1\}$ remains unknown. According to Kacewicz [40], adding and subtracting the integral of $f[A_i(t)]$ in Eq. (20) yields

$$\begin{aligned} u(t_{i+1}) &= u(t_i) + \int_{t_i}^{t_{i+1}} f(u(\tau))d\tau + \int_{t_i}^{t_{i+1}} f(A(\tau))d\tau - \int_{t_i}^{t_{i+1}} f(A_i(\tau))d\tau \\ &= u(t_i) + \int_{t_i}^{t_{i+1}} f(A(\tau))d\tau + \int_{t_i}^{t_{i+1}} [f(u(\tau)) - f(A_i(\tau))]d\tau, \end{aligned} \quad (26)$$

This formulation is exact. Since $A_i(t)$ approximates the exact solution $u(t)$, the third term on the RHS of Eq. (26) becomes negligible and is discarded by Kacewicz [40]. Based on Eq. (22), Kacewicz substitutes $u(t_{i+1})$ and $u(t_i)$ with $y_{i+1}$ and $y_i$, respectively. Consequently, Eq. (26) is simplified to

$$y_{i+1} = y_i + \int_{t_i}^{t_{i+1}} f(A(\tau))d\tau, \qquad t \in [t_i, t_{i+1}] \quad (27)$$

The integral over $I_i$ is divided into the sum of integrals over the sub-subintervals $I_{i,j}$, with $\tau = t_{i,j} + \bar{h}z$, transforming Eq. (27) into

$$y_{i+1} = y_i + N_k \sum_{j=0}^{N_k-1} \frac{\bar{h}}{N_k} \int_0^1 f(A_{i,j}(\tau(z))) dz. \quad (28)$$

As previously mentioned, starting with $y_0 = u_0$, Eqs. (23)-(25) are used to compute the approximate solution $A_{0,j}(t)$ for the sub-subinterval $I_{0,j}$, resulting in $A_0(t)$. Using $A_{0,j}(t)$, the integral on the RHS of Eq. (28) is calculated to yield $y_1$. Repeating this process, $y_1$ is used in Eqs. (23)-(25) to determine the approximate solution $A_{1,j}(t)$ for the sub-subinterval $I_{1,j}$, resulting in $A_1(t)$. Subsequently, $y_2$ is determined according to Eq. (28). By repeating through all the subintervals $I_i$, the algorithm determines the approximate solutions $A_i(t)$ for each subinterval and ultimately constructs the approximate solution $A(t)$ for the ODE system.

In conclusion, the key of the entire algorithm is the computation of the integral in Eq. (28). In Kacewicz's quantum ODE solver, this integral is evaluated using Novak's QIA [41], which is the only step of the algorithm that requires a quantum computer.

**2.3 Quantum Fourier ODE solver**

As discussed in Section 2.1, while the QAEA method can theoretically integrate any continuous function $f(z)$, thereby achieving quantum acceleration [44], it comes with a significant drawback. As highlighted in [43], constructing the circuit of an oracle $\mathcal{R}$ to encode the function $f(z)$ is generally complex. Even in the trivial case of $f(z) = z$, a substantial amount of arithmetic is required to execute the circuit. An important exception is the specific case of $f(z) = \sin^2(mz + c)$ for constants $m$ and $c$. In this scenario, the function can be efficiently encoded using a set of $R_y$ rotation gates, as illustrated in Fig. 3. To address the complexity of general function encoding,

Herbert [43] proposed a quantum integration algorithm called Fourier QMCI by expanding the driving function into a Fourier series. By expressing the driving function as a sum of Fourier series terms, the integration reduces to the computation of trigonometric function-type integrals. This allows the use of the oracle $\mathcal{R}$ illustrated in Fig. 3. Furthermore, Herbert's analysis demonstrated that Fourier QMCI retains the quadratic quantum speedup over classical integration methods. Additionally, it eliminates the need to constrain the driving function to the range [0,1], as required by Novak's QIA, thereby removing the necessity of scaling the driving function.

Based on the idea of Fourier QMCI, we apply a Fourier series expansion to the driving function $f(A_{i,j}(\tau(z)))$ in Eq. (28), expressed as

$$f(A_{i,j}(\tau(z))) = c + \sum_{n=1}^{\infty}[a_n \cos(nwz) + b_n \sin(nwz)] \qquad (29)$$

where

$$\begin{aligned}
a_n &= \frac{2}{T}\int_{-T}^{T} f(A_{i,j}(\tau(z)))\cos\frac{nwz}{2}dz, \quad n = 0,1,2,\ldots \\
b_n &= \frac{2}{T}\int_{-T}^{T} f(A_{i,j}(\tau(z)))\sin\frac{nwz}{2}dz, \quad n = 1,2,3\ldots \\
c &= \frac{a_0}{2}.
\end{aligned} \qquad (30)$$

Here, $w = 2\pi/T$ and $T$ is the period of the periodic piecewise function. According to Eqs. (18) and (23), $f(A_{i,j}(\tau(z)))$ is identified as a polynomial function, allowing its Fourier series coefficients to be readily computed using Eq. (30). However, as previously mentioned, the oracle $\mathcal{R}$ depicted in Fig. 3 is designed to encode the $\sin^2(mz + c)$ function. Thus, we employ the double-angle identities of the trigonometric functions to further transform Eq. (29) into the following expression

$$f(A_{i,j}(\tau(z))) = c + \sum_{n=1}^{\infty}[a_n\cos(nwz) + b_n\sin(nwz)]$$
$$= c + \sum_{n=1}^{\infty}[a_n(1 - 2\sin^2(nwz/2)) + b_n(1 - 2\sin^2(\pi/4 - nwz/2))] \quad (31)$$
$$= c + \sum_{n=1}^{\infty}[-(2a_n\sin^2(nwz/2) + 2b_n\sin^2(\pi/4 - nwz/2)) + a_n + b_n]$$

Substituting Eq. (31) into Eq. (28), we obtain

$$y_{i+1} = y_i + N_k \sum_{j=0}^{N_k-1}\frac{\bar{h}}{N_k}\int_0^1 f(A_{i,j}(\tau(z)))dz$$
$$\approx y_i + N_k\sum_{j=0}^{N_k-1}\frac{\bar{h}}{N_k}\int_0^1\left[c_j + \sum_{n=1}^{N_f}[-(2a_{n,j}\sin^2(nwz/2) + 2b_{n,j}\sin^2(\pi/4 - nwz/2)) + a_{n,j} + b_{n,j}]\right]dz. \quad (32)$$

where $N_f$ represents the truncation order of the Fourier series. Subsequently, we consolidate the identical integral terms in the above equation, yielding

$$y_{i+1} = y_i - \sum_{n=1}^{N_f}\left(2\bar{h}\sum_{j=0}^{N_k-1}a_{n,j}\right)\int_0^1\sin^2(nwz/2)dz$$
$$- \sum_{n=1}^{N_f}\left(2\bar{h}\sum_{j=0}^{N_k-1}b_{n,j}\right)\int_0^1\sin^2(\pi/4 - nwz/2)dz + \bar{h}\sum_{n=1}^{N_f}\sum_{j=0}^{N_k-1}(a_{n,j} + b_{n,j}) + \bar{h}\sum_{j=0}^{N_k-1}c_j. \quad (33)$$

As shown in Eq. (33), by expanding $f(A_{i,j}(\tau(z)))$ using a Fourier series, only $2N_f$ $\sin^2(mz + c)$-type integrals need to be computed to determine $y_{i+1}$. In this work, we refer to this novel quantum ODE solver as the quantum Fourier ODE solver.

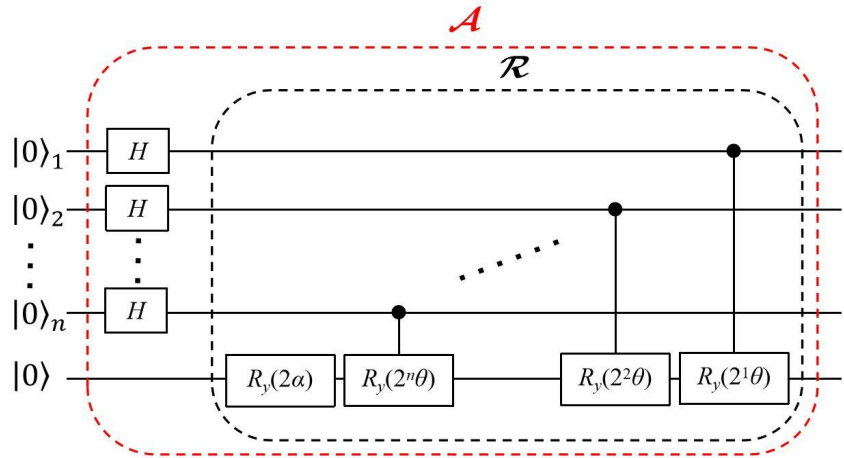

**Fig. 3** Quantum circuit for oracle $\mathcal{A}$ and $\mathcal{R}$ in the quantum Fourier ODE solver.

Finally, we demonstrate the use of QAEA to compute the integral of the $\sin^2(mz + c)$ function, beginning with the following integral

$$I = \int_{b_{min}}^{b_{max}} \sin^2(mz+c)dz. \tag{34}$$

Let

$$b_{max} = b_{min} + (2^n - 1)\Delta, \quad \theta = m\Delta, \quad \alpha = mb_{min} + c,$$
$$z_i = b_{min} + i\Delta, \quad i = 0,1,2,\ldots 2^n\text{-}1, \tag{35}$$

where $\Delta$ is the spacing interval. As discussed in Section 2.1, the quantum circuit depicted in Fig. 3 generates the following quantum state

$$|\psi\rangle = \mathcal{A}|0\rangle_{n+1} = \mathcal{R}(H^{\otimes n} \otimes \mathbf{I})|0\rangle_{n+1} = \mathcal{R}\sum_{i=0}^{2^n-1}\frac{1}{\sqrt{2^n}}|i\rangle_n|0\rangle$$
$$= \sum_{i=0}^{2^n-1}\frac{1}{\sqrt{2^n}}|i\rangle_n \prod_{j=1}^{n} R_y(2^j\theta)^{i_j} R_y(2\alpha)|0\rangle \tag{36}$$
$$= \sum_{i=0}^{2^n-1}\frac{1}{\sqrt{2^n}}|i\rangle_n(\cos(mz_i+c)|0\rangle + \sin(mz_i+c)|1\rangle),$$

where

$$R_y(2^j\theta) = \begin{bmatrix} \cos\dfrac{2^j\theta}{2} & -\sin\dfrac{2^j\theta}{2} \\ \sin\dfrac{2^j\theta}{2} & \cos\dfrac{2^j\theta}{2} \end{bmatrix}, \tag{37}$$

$i_j \in \{0, 1\}$ denotes the $j$-th bit in the binary representation of $i$, with $R_y(2^j\theta)^0$ and $R_y(2^j\theta)^1$ representing the identity matrix $\mathbf{I}$ and $R_y(2^j\theta)$, respectively. From Eqs. (12)-(14), the amplitude of the $|1\rangle$ component of the final qubit in Eq. (36) is

$$\frac{1}{2^n}\sum_{i=0}^{2^n-1}\sin^2(mz_i+c) \approx \frac{1}{b_{max}-b_{min}}\int_{b_{min}}^{b_{max}}\sin^2(mz+c)dz = \bar{I}. \tag{38}$$

By comparing Eq. (34) and Eq. (38), we can obtain the value of the integral in Eq.

(34), i.e., $I = (b_{max} - b_{min})\bar{I}$. However, as discussed in Section 2.1, directly measuring Eq. (36) results in a low probability of obtaining the correct result. Thus, the QAAA is employed to amplify this probability. Fig. 4 illustrates the quantum circuit for QAAA of two qubits, with detailed annotations for the unitary operators described in Eqs. (2) and (16). Finally, the QAEA quantum circuit shown in Fig. 2 is utilized to estimate the amplitude, thereby determining the value of the integral in Eq. (34).

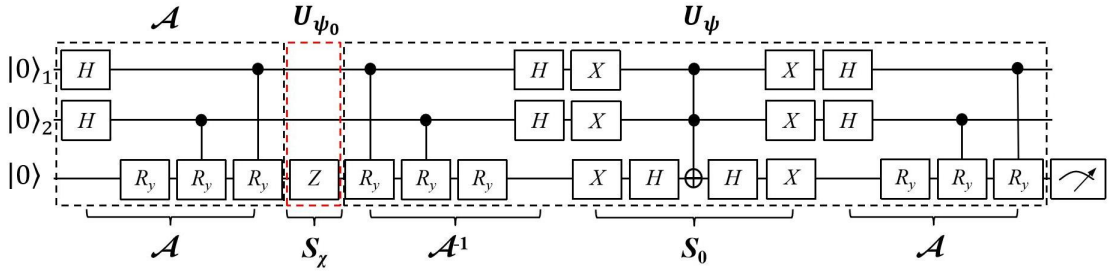

**Fig. 4** Quantum circuit for QAAA of two qubits.

## 3 Numerical Examples

In this section, we apply the proposed quantum Fourier ODE solver to solve various linear and nonlinear PDEs, including the Navier-Stokes (N-S) equations. The accuracy of the algorithm is assessed through the relative $L_2$ error, which is defined as

$$error = \frac{\|u - \hat{u}\|_{L_2}}{\|u\|_{L_2}}, \tag{39}$$

where $u$ and $\hat{u}$ are the analytical/reference solutions and the solutions obtained by the quantum Fourier ODE solver, respectively.

Since the quantum Fourier ODE solver is an explicit algorithm, the duration of the sub-subinterval $\bar{h}$ must satisfy the Courant-Friedrichs-Lewy (CFL) stability condition [46]. For instance, in the case of Burger's equation, the relationship between the local time step $\Delta t_j$ at grid point $j$ and the mesh spacing $\Delta x$ is given by

$$\Delta t_j = C \frac{\Delta x}{u(x_j, t)}, \quad (40)$$

where $u(x_j, t)$ is the velocity, $C$ denotes the CFL number, which must remain less than 1. The numerical simulation operates over the time interval $[0, T]$, and we define $T = N_t \Delta t_{CFL}$, where the global time-step is $\Delta t_{CFL} = \min\{\Delta t_j\}$. As described in Section 2.2, the time interval $[0, T]$ is divided into $n^k$ sub-subintervals of duration $\bar{h}$, thus $T = \bar{h} n^k$. From these two expressions for $T$, we can obtain

$$\frac{\bar{h}}{\Delta t_{CFL}} = \frac{N_t}{n^k}. \quad (41)$$

To ensure that the time partition satisfies the CFL stability condition, we require that the minimum time scale $\bar{h}$ must be less than the CFL time $\Delta t_{CFL}$, i.e.,

$$\frac{\bar{h}}{\Delta t_{CFL}} < 1 \text{ or } \bar{h} < \Delta t_{CFL}, \quad (42)$$

by selecting $n$ and $k$. Furthermore, as discussed in [38-39, 45], the quantum ODE solver requires the error $\varepsilon_1$ in the integral estimate of $f(\cdot)$ to satisfy $\varepsilon_1 = 1/n^{k-1}$. Consequently, $k$ can be determined using the following equation

$$k = 1 + [\ln(1/\varepsilon_1)/\ln n], \quad (43)$$

where $[z]$ is the smallest integer greater than $z$. In this work, we set the error bound $\varepsilon_1 = 0.005$. Thus, $n$ and $k$ can be determined using Eqs. (42) and (43). Notably, to save computational time, we compute the integrals only once in the actual implementation of the algorithm, since the trigonometric function terms in the Fourier series expansion are identical.

## 3.1 Heat conduction equation

In this section, we investigate the properties of the quantum Fourier ODE solver by solving the 2D heat conduction equation. Considering the following equation

$$\begin{cases} u_t = \alpha^2(u_{xx} + u_{yy}), & T \in [0, 0.07], \ \Omega \in [0,1] \times [0,1], \\ u(x, y, 0) = \sin(\pi x)\sin(\pi y), \\ u = 0, \ x \in \partial\Omega. \end{cases} \tag{44}$$

Its exact solution is given by

$$u(x, y, t) = \sin(\pi x)\sin(\pi y)e^{-2\alpha^2\pi^2 t}, \tag{45}$$

where $\alpha^2 = 1$. The second-order central difference scheme is used for spatial discretization. Therefore, the system of ODEs for the 2D heat equation is

$$\frac{\partial u}{\partial t} = \alpha^2 \left( \frac{u_{i+1,j} - 2u_{i,j} + u_{i-1,j}}{\Delta x^2} + \frac{u_{i,j+1} - 2u_{i,j} + u_{i,j-1}}{\Delta y^2} \right). \tag{46}$$

First, we solve Eq. (44) using different mesh sizes: 41 × 41, 61 × 61, 81 × 81, and 101 × 101. In the simulations, the truncation order $N_f$ of the Fourier series is set to 10. To ensure that the CFL number remains below 1, the values of $n$ and $N_k$ are chosen as 256 each. Fig. 5(a) illustrates the relative $L_2$ error of the results obtained by the quantum Fourier ODE solver. It can be seen that the convergence order of the quantum Fourier ODE solver with respect to mesh size is close to the second order, aligning with the theoretical convergence order of the second-order central difference scheme employed. This confirms that the quantum Fourier ODE solver preserves the spatial mesh discretization accuracy.

Next, we investigate the impact of the truncation order $N_f$ of the Fourier series on the performance of the quantum Fourier ODE solver. Fig. 5(b) illustrates the relative $L_2$ error of the results obtained using the quantum Fourier ODE solver for different values of $N_f$. Notably, the mesh size is fixed at 101 × 101, while all other parameters remain consistent with previous configurations. From Fig. 5(b), we observe that increasing the truncation order $N_f$ of the Fourier series does not always reduce the error. This is likely due to the simplicity of the target function, where an excess of higher-order terms in Fourier series can reduce approximation accuracy.

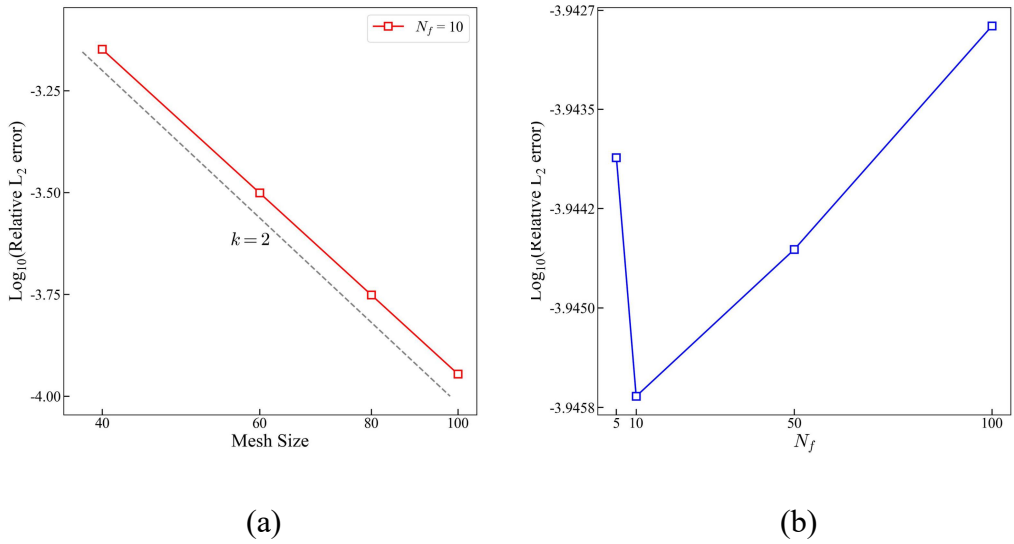

(a)          (b)

**Fig. 5** (a) Convergence order of the relative $L_2$ error versus mesh sizes. (b) Variation of the relative $L_2$ error versus the truncation order $N_f$ of the Fourier series.

Finally, we compare the contour plots of the analytical solution and the solution obtained by the quantum Fourier ODE solver using a mesh with the size of 101 × 101 and $N_f = 10$, as shown in Fig. 6. The results show excellent agreement between the solution obtained by the quantum Fourier ODE solver and the analytical solution. For clearer comparison, Fig. 7 provides the solutions along the vertical line ($x = 0.5$) and the horizontal line ($y = 0.5$). Once again, the two solutions align perfectly. This strong consistency effectively demonstrates the accuracy and validity of the quantum Fourier

ODE solver.

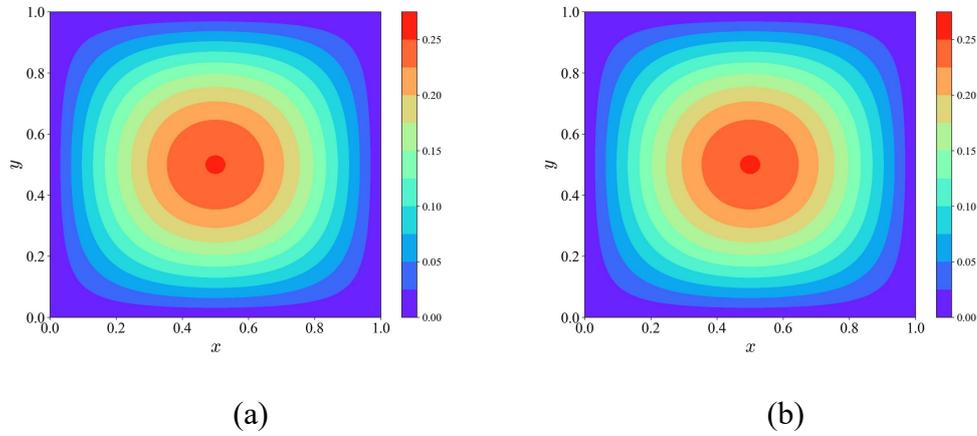

(a)　　　　　　　　　　　　(b)

**Fig. 6** Comparison of (a) the analytical solution and (b) the solution obtained by quantum Fourier ODE solver for the 2D heat conduction equation at $T = 0.07$.

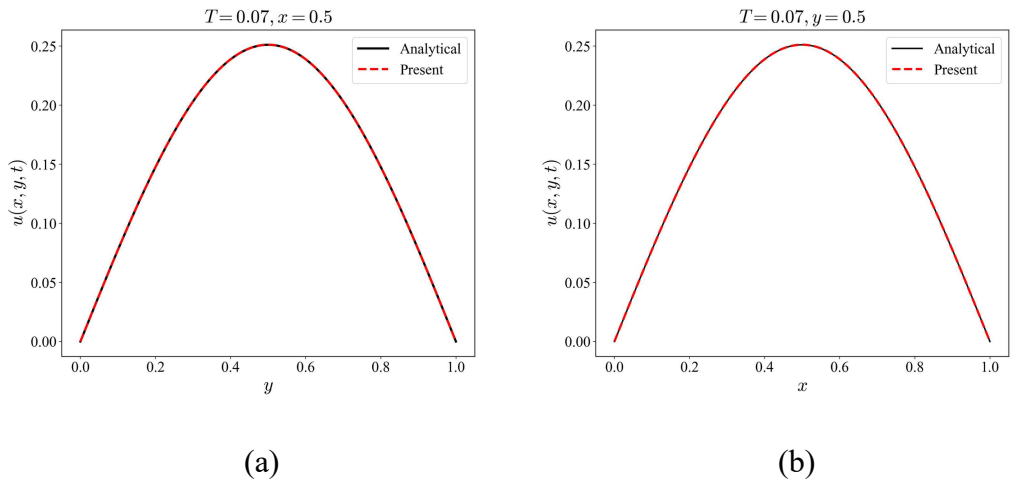

(a)　　　　　　　　　　　　(b)

**Fig. 7** Comparison of results along (a) the vertical line ($x = 0.5$) and (b) the horizontal line ($y = 0.5$) for the 2D viscous Burger's equation at $T = 0.07$.

### 3.2 Viscous Burger's equation

Herein, we generalize our previous study on the heat conduction equation to the more complex 2D viscous Burger's equation, which incorporates both nonlinear convection and diffusion terms. Considering the following equation

$$u_t + uu_x + uu_y = \upsilon(u_{xx} + u_{yy}), \quad \upsilon = 0.01, \quad T \in [0,0.25], \quad \Omega \in [0,1] \times [0,1] \quad (47)$$

Its exact solution is given by

$$u(x,y,t) = \frac{1}{1 + e^{(x+y-t)/2\upsilon}}. \quad (48)$$

and the initial and Dirichlet boundary conditions can be given from the analytical solution. The first-order upwind and second-order central difference schemes are employed to discretize the spatial derivatives of the nonlinear convection and diffusion terms, respectively. The system of ODEs is given by

$$u_t = -\frac{u_{i,j}^2 - u_{i-1,j}^2}{2\Delta x} - \frac{u_{i,j}^2 - u_{i,j-1}^2}{2\Delta y} + \upsilon \left( \frac{u_{i+1,j} - 2u_{i,j} + u_{i-1,j}}{\Delta x^2} + \frac{u_{i,j+1} - 2u_{i,j} + u_{i,j-1}}{\Delta y^2} \right). \quad (49)$$

In the simulation, a uniform mesh with the size of $101 \times 101$ is used. To ensure that the CFL number remains below 1, $n$ and $N_k$ are chosen as 16 and 256, respectively.

The comparison of contour plots between the analytical solution and the solution obtained by the quantum Fourier ODE solver is shown in Fig. 8. It can be seen that the results from the quantum Fourier ODE solver are highly consistent with the analytical solution. To provide more details, Fig. 9 illustrates the solutions of the quantum Fourier ODE solver along the vertical line ($x = 0.5$) and the horizontal line ($y = 0.5$). As shown in Fig. 9, minor discrepancies are observed in regions with sharp gradient changes between the solution of the quantum Fourier ODE solver and the analytical solution. These differences arise due to mesh discretization accuracy and can be minimized by refining the mesh resolution. This case further demonstrates the accuracy and reliability of the quantum Fourier ODE solver.

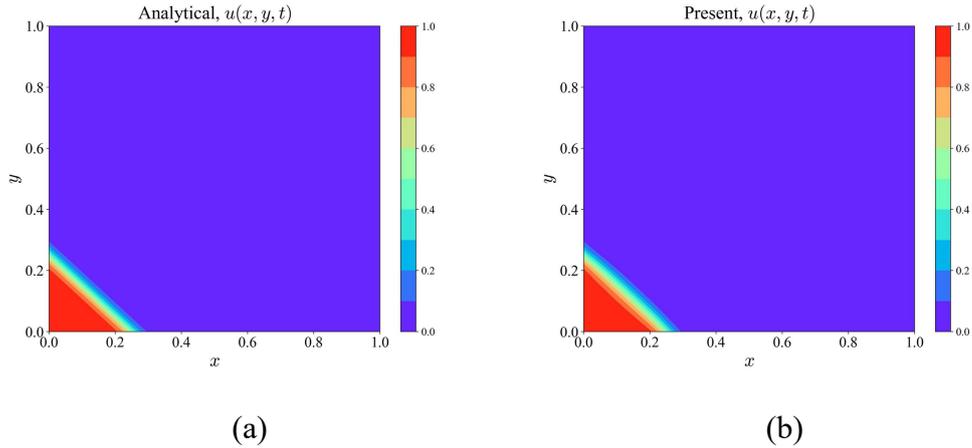

(a)                      (b)

**Fig. 8** Comparison of (a) the analytical solution and (b) the solution obtained by the quantum Fourier ODE solver for the 2D Burger's equation.

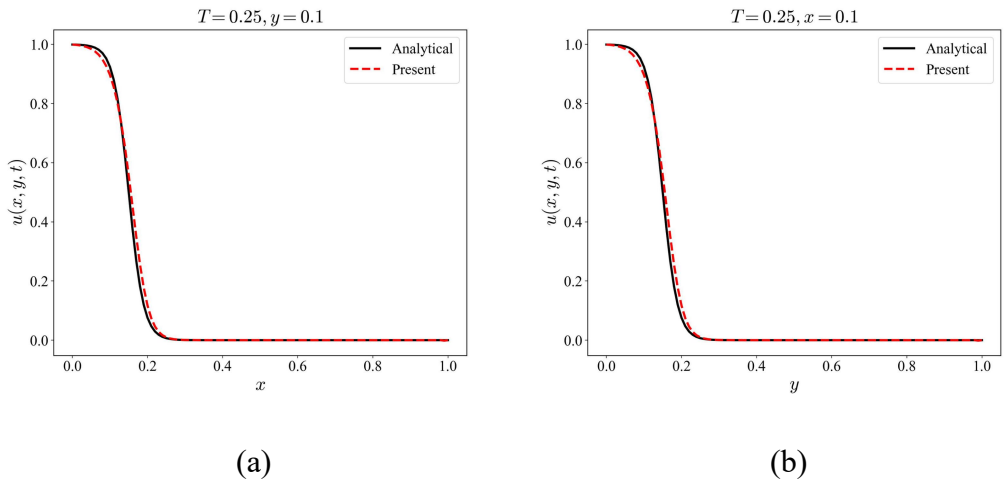

(a)                      (b)

**Fig. 9** Velocity profiles along (a) the horizontal line ($y = 0.1$) and (b) the vertical line ($x = 0.1$) for the 2D viscous Burger's equation at $T = 0.25$.

### 3.3 Coupled viscous Burger's equation

In this case, we solve the more complex 2D coupled viscous Burger's equation, which closely resembles the N-S equations. Considering the following equation

$$\begin{cases} u_t + uu_x + vu_y = \upsilon(u_{xx} + u_{yy}), \\ v_t + uv_x + vv_y = \upsilon(v_{xx} + v_{yy}), \\ \upsilon = 0.01, \quad T \in [0, 0.125], \quad \Omega \in [0,1] \times [0,1]. \end{cases} \quad (50)$$

The exact solution is given by

$$\begin{cases} u(x,y,t) = \dfrac{3}{4} - \dfrac{1}{4 + 4e^{(-4x+4y-t)/32\upsilon}}, \\ v(x,y,t) = \dfrac{3}{4} + \dfrac{1}{4 + 4e^{(-4x+4y-t)/32\upsilon}}. \end{cases} \quad (51)$$

and the initial and Dirichlet boundary conditions can be determined from the analytical solution. Similar to Section 3.2, the first-order upwind and second-order central difference schemes are employed to discretize the spatial derivatives of the nonlinear convection and diffusion terms, respectively. The system of ODEs for the 2D coupled Burger's viscous equation is given by

$$\begin{cases} u_t = -u_{i,j} \dfrac{u_{i,j} - u_{i-1,j}}{2\Delta x} - v_{i,j} \dfrac{u_{i,j} - u_{i,j-1}}{2\Delta y} \\ \quad + \upsilon \left( \dfrac{u_{i+1,j} - 2u_{i,j} + u_{i-1,j}}{\Delta x^2} + \dfrac{u_{i,j+1} - 2u_{i,j} + u_{i,j-1}}{\Delta y^2} \right), \\ v_t = -u_{i,j} \dfrac{v_{i,j} - v_{i-1,j}}{2\Delta x} - v_{i,j} \dfrac{v_{i,j} - v_{i,j-1}}{2\Delta y} \\ \quad + \upsilon \left( \dfrac{v_{i+1,j} - 2v_{i,j} + v_{i-1,j}}{\Delta x^2} + \dfrac{v_{i,j+1} - 2v_{i,j} + v_{i,j-1}}{\Delta y^2} \right) \end{cases} \quad (52)$$

In the simulation, a uniform mesh with the size of $101 \times 101$ is used. To ensure that the CFL number remains below 1, $n$ and $N_k$ are chosen as 16 and 256, respectively.

In Fig. 10, we compare the contour plots of the results obtained by the quantum Fourier ODE solver with the analytical solutions, demonstrating a high degree of consistency. Subsequently, we present the results of the quantum Fourier ODE solver along the horizontal line ($y = 0.5$), as shown in Fig. 11. Similar to Section 3.2, minor discrepancies are observed between the solutions of the quantum Fourier ODE solver and the analytical solutions in regions with sharp gradient changes. However, compared to the results in Section 3.2, the discrepancies between the solutions of the quantum Fourier ODE solver and the analytical solutions in this example are smaller. This is because the solution to Eq. (50) exhibits smaller gradient variations compared to the solution to Eq. (47).

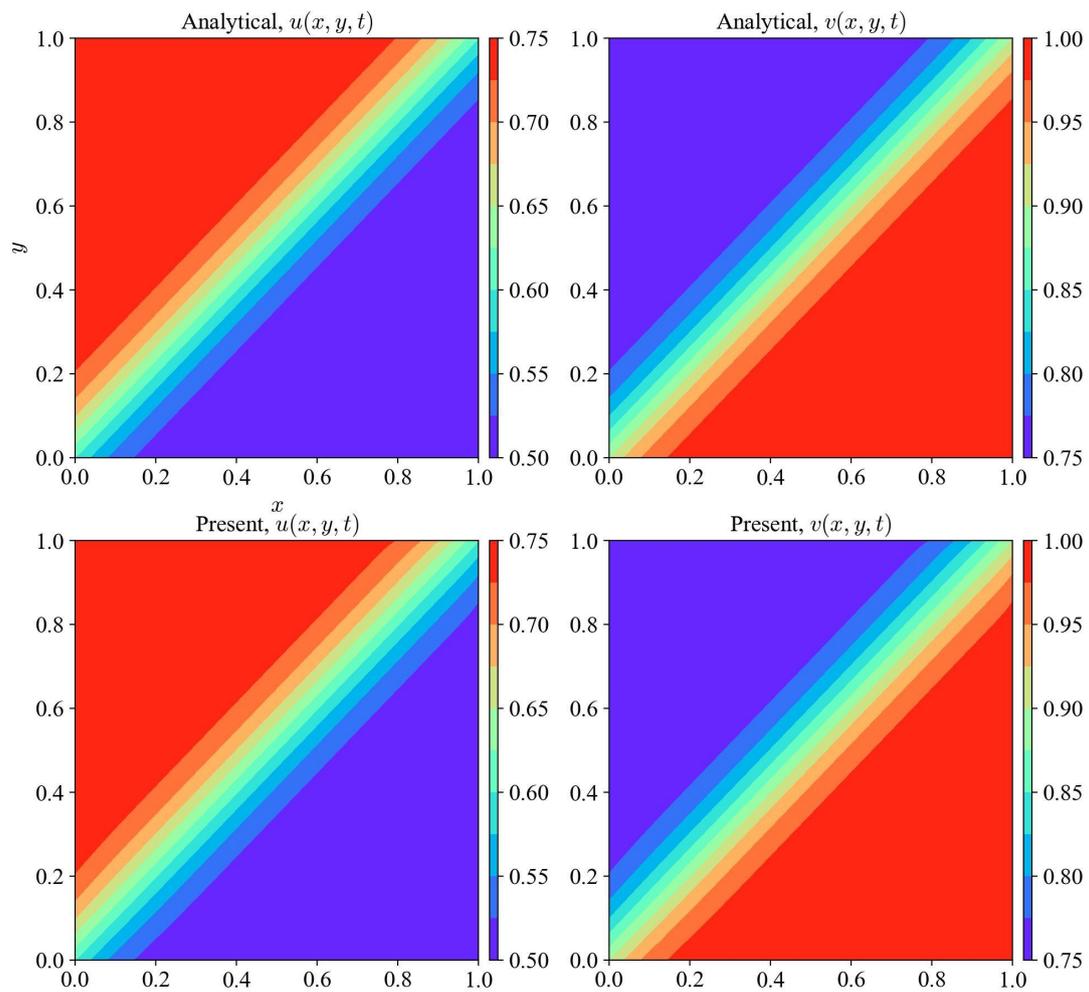

**Fig. 10** Comparison of the analytical solutions and the solutions obtained by quantum Fourier ODE solver for the 2D coupled Burger's equation.

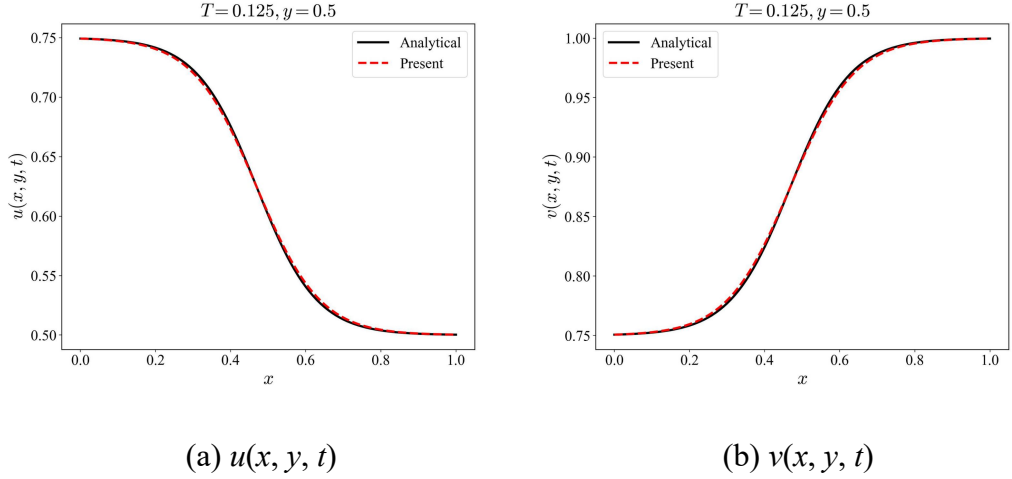

(a) $u(x, y, t)$                  (b) $v(x, y, t)$

**Fig. 11** Velocity profiles along the horizontal line ($y = 0.5$) for the 2D coupled viscous Burger's equation at $T = 0.125$.

### 3.4 Lid-driven cavity flow

In the final case, we simulate the lid-driven cavity flow, which corresponds to the actual N-S equations. Considering the 2D incompressible N-S equations in the stream function-vorticity ($\psi$-$\omega$) formulation

$$\omega_t + u\omega_x + v\omega_y = \frac{1}{Re}(\omega_{xx} + \omega_{yy}), \tag{53}$$
$$\psi_{xx} + \psi_{yy} = -\omega,$$

where

$$u = \psi_y, v = -\psi_x,$$
$$u_{i,j} = \frac{\psi_{i,j+1} - \psi_{i,j-1}}{2\Delta h}, v_{i,j} = \frac{\psi_{i+1,j} - \psi_{i-1,j}}{2\Delta h}, \Delta h = \Delta x = \Delta y. \tag{54}$$

Here, $\psi$, $\omega$, $u$, $v$, and $Re$ are the stream function, vorticity, velocity component in the $x$-direction, velocity component in the $y$-direction, and Reynolds number, respectively. In this case, the spatial domain is specified as $x \in [0, 1]$, $y \in [0, 1]$, and $Re$ is taken as 100. The velocities at the walls follow the no-slip condition, i.e., all walls are stationary except for the top wall, which moves at a constant speed of $u = 1$. The stream function satisfies $\psi = 0$ on all walls, while the boundary conditions for vorticity $\omega$ are

$$\omega_s = -\frac{2(\psi_{s^*} - \psi_s)}{\Delta h^2} \tag{55}$$

on the right, left, and bottom walls, and

$$\omega_s = -\frac{2(\psi_{s^*} - \psi_s + \Delta h)}{\Delta h^2} \tag{56}$$

on the top wall. Here, $s$ denotes the boundary and $s^*$ represents the nodes adjacent to the boundary. The second-order central difference scheme is used to discretize the spatial derivatives. Furthermore, a pseudo-time term is introduced into the second equation of Eq. (53), resulting in the following ODE system

$$\begin{aligned}
\frac{d\omega}{dt} &= -\frac{u_{i,j}(\omega_{i+1,j} - \omega_{i-1,j}) + v_{i,j}(\omega_{i,j+1} - \omega_{i,j-1})}{2\Delta h} \\
&+ \frac{1}{Re}\frac{\omega_{i+1,j} + \omega_{i-1,j} + \omega_{i,j+1} + \omega_{i,j-1} - 4\omega_{i,j}}{\Delta h^2}, \\
\frac{d\psi}{dt} &= \frac{\psi_{i+1,j} + \psi_{i-1,j} + \psi_{i,j+1} + \psi_{i,j-1} - \omega_{i,j}\Delta h^2}{4} - \psi_{i,j},
\end{aligned} \tag{57}$$

In the simulation, a uniform mesh with the size of 41 × 41 is used, and the Fourier series truncation order $N_f$ is set to 5. The subinterval length $T_i$ is set to 0.01, and the number of sub-subintervals is taken as 64.

Fig. 12 presents the stream functions and streamlines obtained by both the CFD solver and the quantum Fourier ODE solver. As seen in Fig. 12, the results from both solvers are largely consistent, effectively capturing the flow structure in this case. To further validate the accuracy of the quantum Fourier ODE solver, we compare the predicted velocity profiles along the horizontal line ($y = 0.5$) and the vertical line ($x =$ 0.5) inside the cavity against reference data, as depicted in Fig. 13. It can be seen that the results from the quantum Fourier ODE solver align closely with the reference data. This case demonstrates the validity and potential of the quantum Fourier ODE solver for solving complex flow problems.

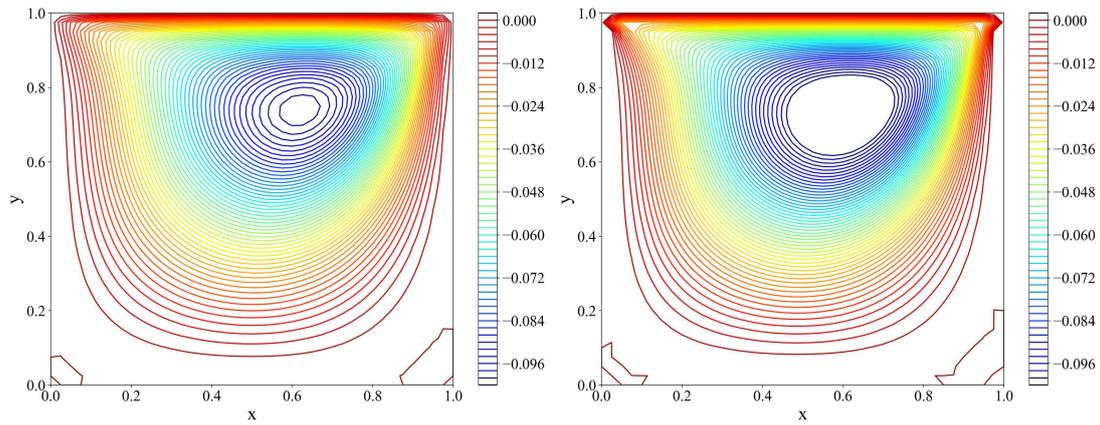

(a) Stream functions. **Left:** CFD solver, **Right:** Quantum Fourier ODE solver.

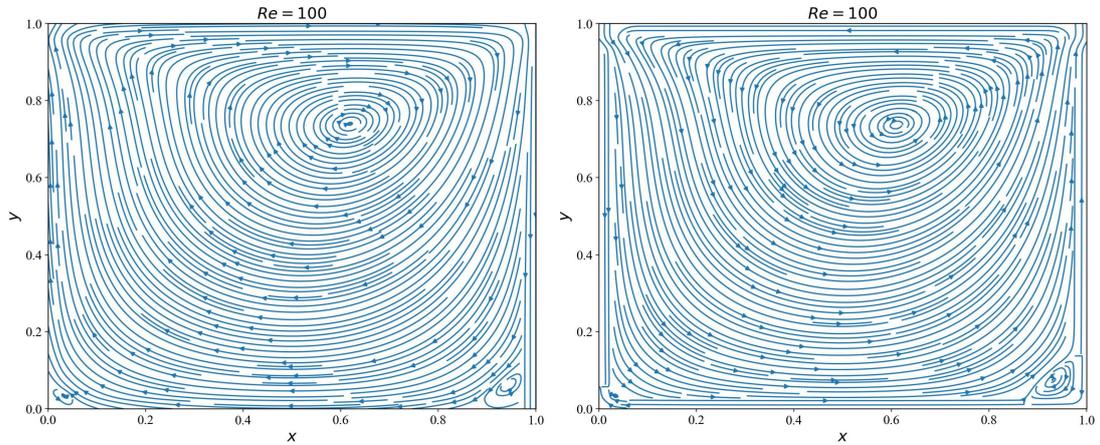

(b) Streamlines. **Left:** CFD solver, **Right:** Quantum Fourier ODE solver.

**Fig. 12** Comparison of (a) stream functions and (b) streamlines obtained by the CFD solver and quantum Fourier ODE solver.

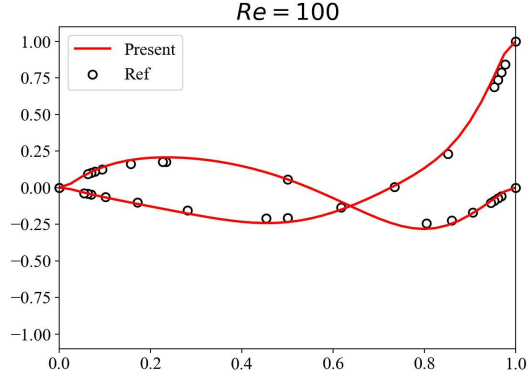

**Fig. 13** Velocity profiles along the horizontal line ($y = 0.5$) and the vertical line ($x = 0.5$) inside the cavity. The black circles denote the reference data.

## 4 Conclusions

In this paper, by introducing the Fourier quantum Monte Carlo integration method to compute the integral of the driving function $f(x)$, we propose a quantum Fourier ODE solver for solving both linear and nonlinear PDEs. Within the framework of the quantum Fourier ODE solver, the complexity of constructing the quantum circuit for the oracle $\mathcal{R}$ that encodes $f(x)$ is significantly reduced by expanding $f(x)$ into a Fourier series, while also avoiding the restriction $f(x) \in [0, 1]$. The performance of the proposed method is evaluated via various linear and nonlinear PDEs, including the N-S equations.

Numerical results indicate that the solutions obtained using the quantum Fourier ODE solver closely match the analytical and reference solutions, with a convergence order relative to mesh size that aligns closely with the theoretical convergence order of the adopted spatial discretization scheme. Furthermore, we observe that increasing the truncation order $N_f$ of the Fourier series does not always improve the accuracy, as an excessive number of high-order terms in Fourier series can degrade approximation.

In conclusion, the quantum Fourier ODE solver provides an efficient approach for solving linear and nonlinear PDEs,. However, the current implementation relies on the traditional QAEA, which requires QPEA with an inverse QFT. The QPEA involves numerous controlled operations and requires a large number of qubits and deep quantum circuits, making it impractical for near-term quantum hardware. Therefore, we plan to implement the quantum Fourier ODE solver using a QAEA that does not require QPEA [47-49] in the future. Furthermore, since QAEA has already been successfully implemented on real quantum computers in some studies [50-51], we aim to explore the deployment of the quantum Fourier ODE solver on actual quantum hardware.

**Acknowledgments**

The research is supported by the National Natural Science Foundation of China (92271103, 12202191) and Research Fund of State Key Laboratory of Mechanics and Control for Aerospace Structures (Nanjing University of Aeronautics and Astronautics) (MCAS-I-0324G04).